\documentclass[aps,prx,twocolumn,floatfix,superscriptaddress]{revtex4-2}

\usepackage{amsmath}  % needed for \tfrac, \bmatrix, etc.
\usepackage{amsfonts} % needed for bold Greek, Fraktur, and blackboard bold
\usepackage{graphicx} % needed for figures
\usepackage{epstopdf}
\usepackage{gensymb}
\usepackage{color}
\usepackage{soul}

\begin{document}
\title{Multistable energy landscapes for adaptive microscopic machines}
\author{Melody Xuan Lim}
 \affiliation{Laboratory of Atomic and Solid-State Physics, Cornell University, Ithaca NY 14853, USA}
 \affiliation{Kavli Institute at Cornell for Nanoscale Science, Cornell University, Ithaca NY 14853, USA}
\author{Zexi Liang}
\affiliation{Laboratory of Atomic and Solid-State Physics, Cornell University, Ithaca NY 14853, USA}
 \affiliation{Kavli Institute at Cornell for Nanoscale Science, Cornell University, Ithaca NY 14853, USA}
\author{Gabriel Alkuino}
\affiliation{Department of Physics, Syracuse University, Syracuse NY 13244, USA}
\affiliation{BioInspired Syracuse: Institute for Material and Living Systems, Syracuse University, Syracuse NY 13244, USA}
\author{Jason Z. Kim}
\affiliation{Laboratory of Atomic and Solid-State Physics, Cornell University, Ithaca NY 14853, USA}
 \affiliation{Kavli Institute at Cornell for Nanoscale Science, Cornell University, Ithaca NY 14853, USA}
\author{Itay Griniasty}
\affiliation{School of Mechanical Engineering, Tel Aviv University, Tel Aviv, Israel}
\affiliation{Center for Physics and Chemistry of Living Systems, Tel Aviv University, Tel Aviv, Israel}
\author{Teng Zhang}
\affiliation{Department of Mechanical and Aerospace Engineering, Syracuse University, Syracuse NY 13244, USA}
\affiliation{BioInspired Syracuse: Institute for Material and Living Systems, Syracuse University, Syracuse NY 13244, USA}
\author{Paul L. McEuen}
\affiliation{Laboratory of Atomic and Solid-State Physics, Cornell University, Ithaca NY 14853, USA}
 \affiliation{Kavli Institute at Cornell for Nanoscale Science, Cornell University, Ithaca NY 14853, USA}
\author{Itai Cohen}
 \affiliation{Laboratory of Atomic and Solid-State Physics, Cornell University, Ithaca NY 14853, USA}
 \affiliation{Kavli Institute at Cornell for Nanoscale Science, Cornell University, Ithaca NY 14853, USA}
\begin{abstract}
The past few years have seen great strides in our ability to build synthetic microscopic machines. However, the function of such machines is often controlled directly by externally applied fields that deterministically specify the instantaneous machine dynamics. A crucial step towards machines that can respond adaptively to changes in their environment is the ability to program multiple functions that actuate under the same external driving field, so that their internal state dictates which function is executed. 
Here, we demonstrate that energy landscapes with designed multistability enable the same externally applied field to drive multiple configurations and dynamic responses in microscopic machines, enabling increasing levels of autonomy. We show three examples. First, we  write a bistable energy landscape into a microscopic device, enabling the device to exhibit two stable mechanical configurations under the same external magnetic field. Next, adding a second degree of freedom enables differing dynamic responses to the same external magnetic field, which we direct into net displacement of the environment. Finally, we demonstrate how a microscopic machine with a continuous symmetry autonomously channels a single degree-of-freedom magnetic actuation into locomotion and adaptively responds to forces induced by other machines. 
\end{abstract}
\maketitle

The rational design of microscopic machines that transduce external energetic inputs to do work on their environment is a burgeoning area of science with broad implications for microrobotics. Recent studies have focused on synthetic micro-machines powered by external energy sources, such as light~\cite{palagi2016structured,miskin2020electronically,reynolds2022microscopic,hong2024optoelectronically,hanson2025electrokinetic}, acoustic fields~\cite{ren20193d,deng2023acoustically,mahkam2025multiorifice,shi2025ultrasound}, or time-varying magnetic fields~\cite{tasci2016surface,yang2020reconfigurable,hu2021magnetic,liu2022creating,gu2022artificial,smart2024magnetically,abbasi2024autonomous}. Magnetically actuated microscopic machines have received particular attention due to their potential for highly programmable remote actuation. The conformational changes of these micro-machines, however, are often specified by the externally applied magnetic field, rather than adaptively determined at the device level. Such global control prevents differential functionality with respect to changes in the device environment, and limits applications to situations where there can be close feedback between the current machine state and externally applied fields. A crucial step towards enabling adaptive differential responses in such machines would entail building devices that can store information about past environmental stimuli in an internal state, and differentially actuate on the basis of both the externally applied field and their internal state.

\begin{figure*}
    \centering
    \includegraphics[width = 1.3\columnwidth]{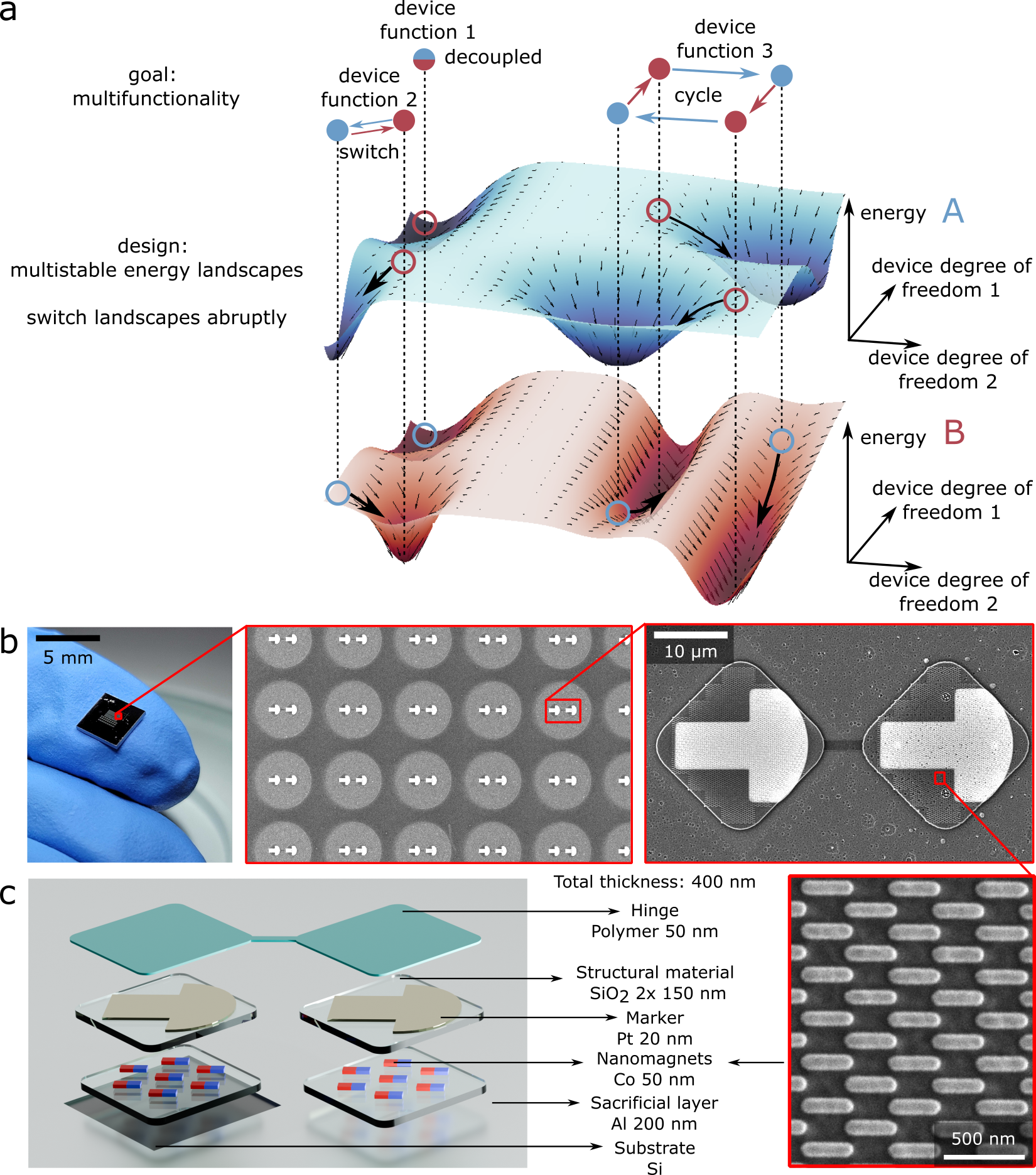}
    \caption{\textbf{Fabricating devices that can differentially respond to the same external field using programmable energy landscapes.} (a) (Top) Configuration space of a single device. A single sequence of changes in the external field could either produce no change (function 1), switch between two states (function 2), or a cycle between four states (function 3). We design such multifunctionality by designing the energy landscape (colored surfaces) at each external field, together with its gradients (black arrows). Possible conformations at each value of the external field are encoded as potential minima (dashed lines). Changing the external field drives conformational changes. (b) Massively parallel fabrication of magneto-elastic devices with programmable energy landscapes. (Left) Photograph of a 5 mm square silicon chip containing several hundred devices. (Center) Scanning electron micrograph of an array of devices. (Right) Scanning electron micrograph of a single device.  (c) (Left) Perspective drawing showing the layer structure of a device. The panel on the left is fabricated directly on the Si substrate and will remain bonded to the substrate, while the panel on the right is fabricated on the sacrificial Al layer and after release remains tethered only through the polymer hinge. (Right) Scanning electron micrograph of a nanomagnet array.}
    \label{fig:fabrication}
\end{figure*}

In biology, proteins encode internal states in their conformation~\cite{movassagh2010nucleotide,catterall2020conformational,kim2021functional,praetorius2023design}, which determines what biological function is produced when ATP is metabolized. Inspired by this approach, here, we investigate how multistable energy landscapes can be used to encode internal device states and direct adaptive actuation in synthetic magneto elastic microscopic machines (Fig.~\ref{fig:fabrication}a). Our thesis is that multistability enables the same externally applied magnetic field to address different device functions (groups of colored dots in configuration space, top of figure). When the applied field is changed, depending on its initial configuration, a device could then enact no response at all (function 1 in Fig.~\ref{fig:fabrication}a), switch between two conformational states (function 2 in Fig.~\ref{fig:fabrication}a), or cycle between conformational states to produce net displacements in the device's local environment (function 3 in Fig.~\ref{fig:fabrication}a). The possible device configurations at each value of the external field are encoded as potential minima in the energy landscapes. When the external field is abruptly switched from one value to the other, the stable device conformations also switch. The internal device structure then drives conformational changes between stable states, which appear as gradients on the energy landscape (black arrows on colored surfaces in Fig.~\ref{fig:fabrication}a). Thus, these functions and the device response to the external fields can be programmed by designing the energy landscapes for each value of the applied field (colored surfaces Fig.~\ref{fig:fabrication}a). 
% Since the initial device configuration depends on the history of external forces applied to the device, the device energy landscapes encode an internal state that determines which of its possible functions is executed under the action of a changing external field. 

We explore this strategy by designing and fabricating microscopic machines whose mechanics are governed by magneto-elastic energy landscapes~\cite{jang2015undulatory,kim2018printing,cui_nanomagnetic_2019,xu2019millimeter,gu2020magnetic,zhang2021voxelated,dong2022untethered, smart2024magnetically,ren2024design}. Magneto-elastic platforms have several advantages. Magnetic interactions are easy to program since there are no mobile charges to provide screening effects. By combining these magnetic interactions with elastic forces, it is possible to design and construct energy landscapes with many potential minima. 
% Additionally, transitions between these constructed energy minima can be actuated by applying directed torques in the form of externally applied magnetic fields. 

We use this platform to demonstrate three devices with increasing levels of autonomy from the driving magnetic field. First, we show a two-state microscopic bistable device, where two device states, depending on the history of actuation, are possible over a range of field strengths. We then illustrate how the addition of a second conformational degree of freedom enables a device that exhibits different conformational changes in response to the same destabilizing magnetic driving. These dynamic responses can be pieced together to generate a net displacement of the device's fluidic environment. Finally, we expand on this paradigm to demonstrate energy landscapes that enable microscopic machines to independently ``metabolize" a neutral energy input, produce autonomous locomotion, while sensing and interacting with other machines. 

\begin{figure*}
    \centering
    \includegraphics[width = 1.3\columnwidth]{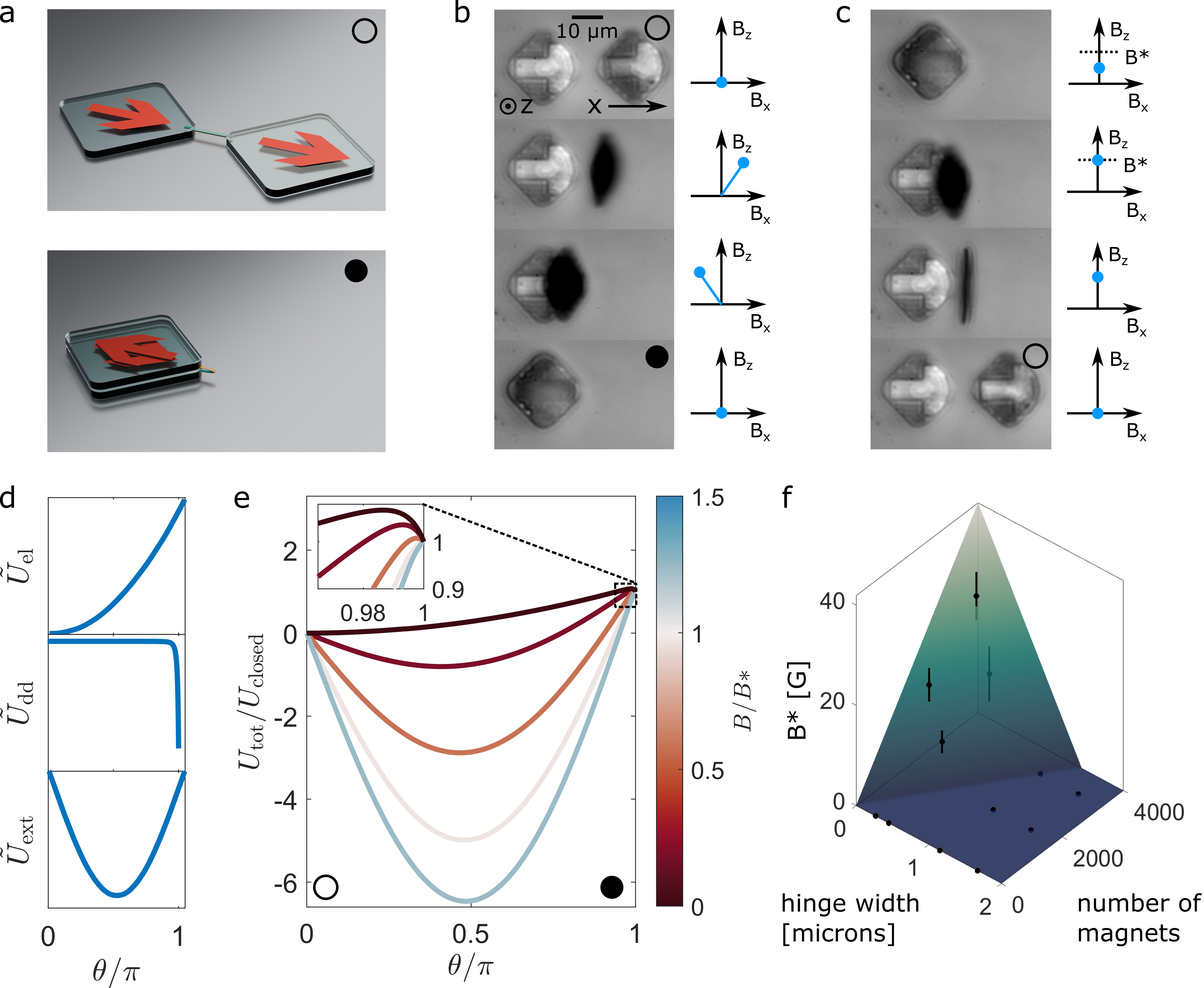}
    \caption{\textbf{Encoding bistability in a microscopic device to enable multiple configurations under the same external field.} (a) Perspective drawings of the two stable conformations. (b) Sequence of optical microscope images and accompanying magnetic field magnitudes while closing a two-state device. For sufficiently large number of magnets on the fixed panel and sufficiently narrow hinge width, magnetic interactions stabilize the closed position when the magnetic field is turned off. (c) Sequence of images showing the device re-opening upon the application of a finite magnetic field~$B^*$. (d) Non-dimensionalized components of the energy landscape for the actuated two-panel magnetic device, from top to bottom: elastic energy~$\tilde{U}_\mathrm{el}$, magnetic dipole-dipole interaction energy~$\tilde{U}_\mathrm{dd}$, dipole-field interaction energy~$\tilde{U}_\mathrm{ext}$ for a magnetic field in the z-direction. (e) Total energy~${U}_\mathrm{tot}$ normalized by the energy of the closed state~${U}_\mathrm{close}$, for varying values of the vertical external magnetic field~$B$ normalized by the critical magnetic 00
    -field~$B^*$. (f) Plot of the experimentally measured~$B^*$ as a function of the hinge width and the number of magnets on the fixed panel (black data points), together with an analytical prediction based on bifurcation theory (colored surface). Error bars on the data points correspond to the standard error associated with device variation. }
    \label{fig:twostate}
\end{figure*}

To create microscale magneto-elastic devices with designable energy landscapes, we require precise control over magnetic dipole-dipole interactions within a device, flexible elements with controlled elasticity, and the coupling of the device to an external magnetic field. We build on recent developments in micro-robotics~\cite{cui_nanomagnetic_2019,dorsey2019atomic,smart2024magnetically,liang2025magnetic} to precisely control via lithography the position, orientation, and density of magnetic dipoles within microscopic rigid panels, which we then connect with flexible hinges (Fig.~\ref{fig:fabrication}b,c; see Supplementary Information for further details). The net dipole moments on each panel ranged from 0 to $3.3\times10^{-12}$ Am$^2$. The flexible hinge was comprised of a DUV42P anti-reflective coating that was 50 nm thick by 5.4 \textmu m long with hinge widths ranging from 270 nm to 1670 nm, and a modulus of 1 GPa (see Supplementary Information for further details on device parameters). For a given device conformation, the elastic energy landscape is then specified by the geometrical parameters of the hinge, while the magnetic energy landscape is controlled by the relative position and orientation of the panel dipole moments. 

We first demonstrate that bistability enables two possible configurations of the device under a range of applied magnetic fields. We fabricated bistable devices consisting of two panels connected by an elastic hinge. Each microscopic (20 \textmu m long) panel was covered with a set number of distributed magnetic dipoles facing in the same direction. One of the panels was free to move in response to an external magnetic field while the other was fixed. We first consider the open and closed states shown in Figure~\ref{fig:twostate}a: one where both panels lie flat on the surface and the elastic energy is minimized (open circle, referred to as ``open"), and one where the free panel is stacked on top of the fixed panel, and the magnetic interaction energy is minimized (filled circle, referred to as ``closed"). Since both states are possible at zero applied field, bistability decouples the device configuration from the instantaneous magnetic field. Instead, the device configuration and dynamic response depends on its history of actuation. 

To elucidate how bistability can decouple a device's response from the driving, we consider how the open and closed states behave under the same applied magnetic fields. Here, we performed experiments on a device with magnetic dipole moments of $3.3\times10^{-12}$ Am$^2$ on each panel and a hinge width of 280 nm. In the open state, the free panel can simply align with the direction of the applied magnetic field (Fig.~\ref{fig:twostate}b). This alignment allows for closing the device by bringing the free panel close to the fixed panel and allowing for their magnetic interaction to stabilize their configuration. Once closed, the device is insensitive to magnetic fields below a threshold field amplitude $B^*=29\pm5$ G (Fig.~\ref{fig:twostate}c). Above this threshold, the closed state destabilizes and the panel is once again free to align with the applied field orientation. This differential actuation demonstrates how an energy landscape can produce history-dependent device motion in microscopic machines. Although here we generated the device history via the action of a changing magnetic field, any local force that acts to deflect the free panel into the closed state could produce the same effect, allowing for the device configuration to store a memory of past forces applied to the free panel. 

\begin{figure*}
    \centering
    \includegraphics[width = 1.5\columnwidth]{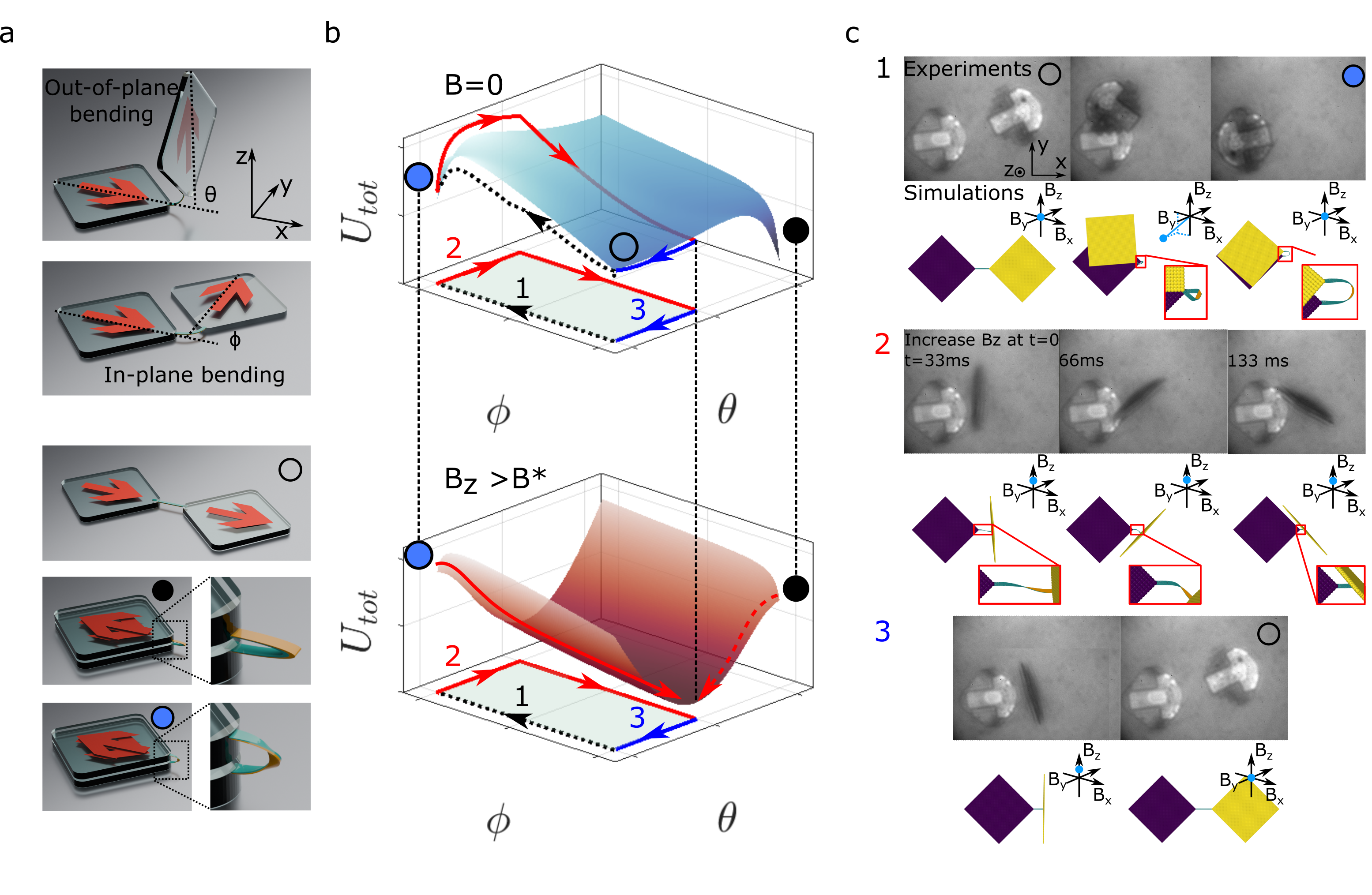}
    \caption{\textbf{Using internal device states to enable multiple conformational changes under the same external field.} (a) Illustration of the two degrees of freedom of the system~$\theta$ and~$\phi$, together with the three stable states of the system, labeled by symbols (open circle, black circle, blue circle for open state, closed state, and twist-closed state respectively). (b)  Analytical energy landscapes for the three-states illustrated in (a), for two different magnetic field conditions. (Top) Total energy,~$U_\mathrm{tot}$ at zero magnetic field, consisting of the sum of the elastic energy and dipole-dipole interaction energy. Colored lines on the energy landscape show the energy of the simulated work cycle. The shaded area in configuration space corresponds to the net displacement of the environment due to the cycle. (Bottom) $U_\mathrm{tot}$ under an out-of-plane magnetic field~$B_z$ greater than the critical re-opening field~$B^*$, consisting of the sum of the elastic energy, dipole-dipole interaction energy, and the magnetic potential energy due to the external field. Red (dotted) lines show the conformational change executed by a device starting in the twist-closed (closed) state when the external field is switched abruptly from~$B=0$ to~$B_z>B^*$. (c) Series of optical microscope images showing the two-panel device executing a work cycle, marked by symbols and numbers corresponding to the stages of the work cycle illustrated in (b), paired with frames from a molecular-dynamics simulation. Stage 1 (dotted black line in (b)): the free panel is manipulated using an external magnetic field to lie on top of the fixed panel. Stage 2 (red line in (b)): when the magnetic field is switched to point out of the plane, the free panel spontaneously untwists. Stage 3 (blue line in (b)): When the external field is switched off, the device returns to the open state. }
    \label{fig:threestate}
\end{figure*}

In this device, the threshold field $B^*$ is determined by the balance between the elastic and magnetic device energies. The total energy of this device consists of the sum of three components: the hinge elastic energy, the magnetic dipole-dipole interaction energy, and the interaction energy of the free panel with the external magnetic field (Fig.~\ref{fig:twostate}d. Gravitational potential energy is negligible, see Supplementary Information for estimates of its magnitude). At zero magnetic field, the open and closed states correspond to minima in the energy landscape when the bending angle, $\theta$, is equal to 0 and $\pi$ respectively (dark red curve in Fig.~\ref{fig:twostate}e). As the external magnetic field is increased, for example in the $z$ direction, the angular location of the open state moves to align with the applied magnetic field. For small fields, the closed state remains stable at the same configuration. The energy landscape undergoes a bifurcation at $B=B^*$, where the closed state loses stability. For $B > B^*$ the only stable configuration for the free panel corresponds to being aligned with the external magnetic field (Fig.~\ref{fig:twostate}e). We plot the measured $B^*$ (black data points) versus the hinge width and number of magnets on the fixed panel in Fig.~\ref{fig:twostate}f. We find excellent agreement between the measured values and predictions from bifurcation theory (colored surface in Fig.~\ref{fig:twostate}f). This two-panel, one elastic degree-of-freedom microscale device thus provides a minimal example of using energy landscapes to decouple from the applied magnetic field, with the range of decoupled motion determined by programmed device parameters. Such a two-state device could be used to detect small changes in magnetic field when operated close to~$B^*$. 

Next, we demonstrate that a device with two conformational degrees of freedom can execute orthogonal conformational changes in response to the same applied magnetic field. As an example, we use the same two-panel microscale device, and expand its configuration space to include twisting by an angle $\phi$ about the~$z$ axis. Specifically, twisting the free panel about the~$z$-axis and anchoring it on top of the fixed panel produces a third stable state at zero applied magnetic field (in addition to open and closed), which we refer to as ``twist-closed" (blue filled circle, device configuration illustrated in Fig.~\ref{fig:threestate}a). This state has the same dipole-dipole interaction energy as the closed state, but the elastic hinge has been twisted by 180 degrees, increasing the elastic energy compared to the closed state and reducing its threshold magnetic field (top panel Fig.~\ref{fig:threestate}b, see Supplementary Information for details on the analytical calculations). 

When the external magnetic field is switched to a sufficiently large out-of-plane value, this twisted hinge acts as an internal mechanism to drive conformational changes (energy landscape shown in bottom panel of Fig.~\ref{fig:threestate}b). As previously discussed, when an out-of-plane magnetic field~$B>B^*$ is applied to the closed state, the free panel moves to align with the field direction~$\theta=\pi/2$, tracing out a one-dimensional line along the~$\theta$ axis in Fig.~\ref{fig:threestate}b (dashed red line in bottom panel). In contrast, applying the same magnetic field to the twist-closed state results in a complex sequence of deformations where the free panel first lifts to align with the field direction, then spontaneously releases the twist in the hinge, corresponding to decreasing~$\phi$ (solid red arrow in Fig.~\ref{fig:threestate}b, images in Fig.~\ref{fig:threestate}c, middle row), while remaining aligned with the vertical external field. These dynamics can be understood by examining the energy landscape (Fig.~\ref{fig:threestate}b, bottom panel), which shows that a device in the twist-closed state first experiences strong gradients to reduce~$\theta$, before the shallower gradients in the ~$\phi$ direction act to untwist the hinge. This separation of scales is due to the different elastic moduli governing the twist and bend degrees of freedom, i.e. the hinge geometry. The closed and twist-closed states therefore respond to an identical magnetic perturbation with conformational changes that are orthogonal in the device configuration space.  

Switching between the two energy landscapes illustrated in Fig.~\ref{fig:threestate}b allows us to combine the dynamic responses of the device to generate an area-enclosing cycle in configuration space, and therefore generate a net displacement in the device environment~\cite{purcell2014life,lauga2011life,shapere1989geometry}. For instance, beginning from the open state, we guide the device into the twist-closed state by applying a sequence of magnetic torques to increase~$\phi$ (dashed black line in Fig.~\ref{fig:threestate}b, row 1 in Fig.~\ref{fig:threestate}c), storing the magnetic field rotation as internal elastic energy of the twisted hinge. Next, we switch to the second energy landscape by applying an out-of-plane field, which lifts the free panel and releases the stored elastic energy into spontaneous untwisting. This untwisting motion brings the hinge back to its flat configuration (solid red line, row 2 in Fig.~\ref{fig:threestate}c). Finally, we remove the magnetic field, switching back to the zero-field energy landscape and completing the cycle with a return to the open state (solid blue line, row 3 in Fig.~\ref{fig:threestate}c). Illustrating the actuation sequence on the device configuration space (Fig.~\ref{fig:threestate}b) reveals the area enclosed by the cycle of shape changes, corresponding to the net displacement on the fluid environment due to the action of the microscopic machine. 

In order to validate the illustrated configurational cycle, we perform molecular-dynamics simulations to fully resolve the hinge conformation during the work cycle (Fig.~\ref{fig:threestate}c, see Supplementary Information for simulation details). The simulations closely reproduce the observed experimental device conformations, including the dynamics of untwisting during the transition out of the twist-closed state (row 2 of Fig.~\ref{fig:threestate}c). Furthermore, plotting the sum of the calculated elastic and magnetic dipole-dipole interaction energies reveals close agreement between the analytically calculated energy landscape (colored surface) and the simulated trajectory (black, red, and blue lines) through the energy landscape. 
The agreement between theoretical calculation, simulation, and experiment highlights that the key ingredients for designing and constructing this microscopic machine are the magnetic and elastic potential energy landscapes, which specify different dynamic responses to the same external perturbation. 

\begin{figure*}
    \centering
    \includegraphics[width = 1.5\columnwidth]{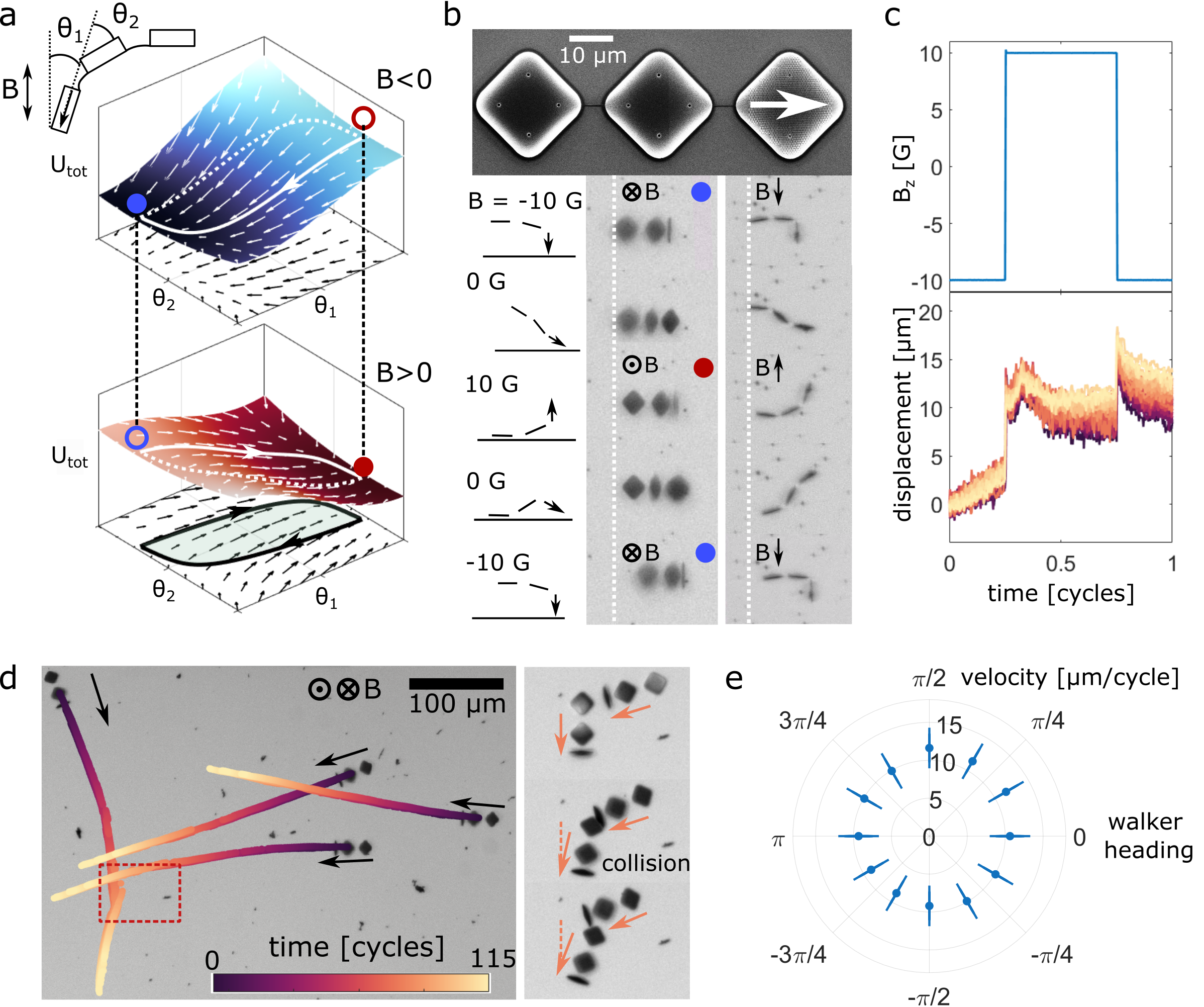}
    \caption{\textbf{Using symmetric energy landscapes for  locomotion.} (a) (Top left) Schematic of a three-panel device with two degrees of freedom,~$\theta_1$ and~$\theta_2$. The device configuration is symmetric with respect to rotations about the z-axis. (Main panels) Two-dimensional projection of the energy landscape of the three-panel device for two different values of the external magnetic field. The energy landscape here consists of the sum of the elastic energy and the magnetic potential energy due to the external field. For each value of the external magnetic field, there is a single energetic minimum configuration in the energy landscape (left side, minima marked by blue dot and red dot). Gradients in the energy landscape (arrows) between the two states form a nonreciprocal cycle (projection shown on bottom row), with enclosed area (shaded area in projection) corresponding to the net displacement of the device. (b) (Top) Scanning electron microscope image of the structure of a magnetic walker, consisting of a magnetic ``head" and two non-magnetic ``tail" panels connected by elastic hinges. (Bottom) Time-series images and illustrations of a single step in the work cycle. (Left) Schematic of the conformation of the three panels throughout the cycle. (Center) Optical microscope images of a walker undergoing a single step in the work cycle, where the magnetic field oscillates in- and out-of-plane. (Right) Optical microscope images of a walker undergoing a single step in the work cycle, where the magnetic field oscillates in-plane, along the~$\pm y$ direction. Dotted lines highlight the initial position of the magnetic ``tail". (c) Magnetic field during a single cycle, with synchronized walker displacement. (d) Center-of-mass motion of four magnetic walkers with tails facing several directions, actuated at 25 Hz, with the time indicated by the shared color scale. Two walkers collide in the region highlighted with a red box, which is shown in a time-series of images on the right. (e) Walker displacement per cycle as a function of the walker heading direction, for actuation frequency 0.1 Hz. }
    \label{fig:kinesin}
\end{figure*}

In order to further increase device autonomy, we envision designing a device which can channel a magnetic perturbation into conformational changes in any direction based on its state history. For example, we illustrate such a device in Fig.~\ref{fig:kinesin}a, which is symmetric with respect to rotations about the vertical axis. As a result, actuation with a vertical magnetic field autonomously drives configurational changes in any of the possible in-plane device orientations. If these configurational changes are non-reciprocal over a magnetic field cycle, this device could autonomously perform work cycles whose output depends only on the initial device orientation. 

Actuating a work cycle via a magnetic field that varies only about the vertical symmetry axis, i.e. a one degree-of-freedom magnetic field input, requires a further extension of the design paradigm illustrated in Fig.~\ref{fig:threestate}. For example, we consider a device consisting of a magnetic head connected to two undecorated panels via elastic hinges (Fig.~\ref{fig:kinesin}a). Under two different magnetic fields (labeled here as~$B>0$ and~$B<0$), the energy landscape of the device has two different potential minima (red and blue points). If the magnetic field is switched continuously and adiabatically between these two extremes, the device conformation will trace the one-dimensional line that connects the two minima, thus producing no work. As with the two-panel device, however, we can take advantage of the device dynamics and the gradients on the energy landscape to generate non-reciprocal cycles (arrows in Fig.~\ref{fig:kinesin}a). The gradients specify the trajectory of the device through configuration space from a given initial condition, and do not have to be reciprocal with respect to changes in the magnetic actuation due to relaxation times associated with the internal device dynamics (e.g. hinges, panel motions, etc.). As a result, when the magnetic field is abruptly switched between the two conditions, the conformational change of a device is specified by the gradients in the device energy landscape between two minima, and for certain ranges of switching frequency can produce work cycles (Fig.~\ref{fig:kinesin}a, right side, see Supplementary Information for further details on actuation frequency).

Experimentally, we realize such an actuation scheme using the three-panel device shown in Fig.~\ref{fig:kinesin}b. The three panels are connected by two hinges with width 330 nm and thickness 50 nm, with the first panel having a net dipole moment of~$3.3\times10^{-12}$ Am$^2$, while the second and third panels have no magnetic moment. All three panels are free to move in response to external forces. Under the action of an external magnetic field oriented in- and out-of-the plane, the magnetic head particle experiences a torque that aligns it with the current magnetic field direction. This magnetic torque then propagates elastically to the two tail particles. This differential torque, together with the fluid-solid interaction, produces a characteristic timescale for the structure that enables the use of transient dynamics on the energy landscape, as illustrated in Fig.~\ref{fig:kinesin}a. As a result, most of the productive locomotion is initiated as the magnetic field switches between the two field conditions (Fig.~\ref{fig:kinesin}c), i.e. when the magnetic torque drives the most rapid reorientation of the head.

This swimming motion utilizes the previously demonstrated beating action of artificial cilia and flagella~\cite{dreyfus2005microscopic,khalil2014magnetosperm,jang2015undulatory,li2016magnetically,hanasoge2018microfluidic,yang2020reconfigurable,islam2021highly,wang2022cilia,zhang20233d} (see Supplementary Information for simulations of the induced fluid flow). Here, however, the external magnetic field does not define an actuation direction. Each walker is thus free to utilize the external field to power locomotion in a direction set by the direction of the tail relative to the head (in turn determined by its state history), rather than a single global signal (Fig.~\ref{fig:kinesin}d). As a result, swimmers are able to autonomously navigate in any direction on the two-dimensional plane with equal efficiency (Fig.~\ref{fig:kinesin}e). Furthermore, this in-plane autonomy enables walkers to adaptively respond to forces exerted by other walkers, here either steric contact forces or hydrodynamic forces, and alter their navigational heading in response (example highlighted in Fig.~\ref{fig:kinesin}d).   

We have demonstrated a minimal model system for the design and programming of multi-stable energy landscapes and their gradients, which can be harnessed to perform autonomous microscale work both through conformational changes and by locomotion. Our magneto-elastic energy landscapes are directly constructed from the addition of their base components, and are straightforward to compute from the device configuration, allowing for direct simulation and analytical calculation to directly inform the design process. The ability to precisely design energy landscapes represents a key step towards testing and realizing recently proposed theoretical design principles for microscale machines~\cite{yang2023bifurcation,chatzittofi2025mechanistic}.  

 Our work demonstrates that the challenge of autonomous differential actuation in microscale robotics can be framed in the lens of designing features of a multistable energy landscape. Although here we applied magnetic forces to switch device configurations, our approach extends to any external force that could change the device configuration, including mechanical, frictional, fluidic, optical, or electronic signals. Importantly, these cues do not have to be addressed via a global control framework, but can be embedded in the local environment of a device, or even generated collectively by the devices themselves. We envision extending our work to produce devices that sense their environment, encode changes in their conformational state, and autonomously adapt their response to their local environment. 

\subsection*{Experimental procedures}
To perform an experiment, we first magnetize the particles using a strong electromagnet (field $>$100 mT during magnetization). At this point, each sample consists of a piece of silicon, covered with an aluminium release layer, with devices on top of the release layer. This sample is then submerged in a solution of 0.125\% tetramethylammonium hydroxide (TMAH) in water. The TMAH undercuts the aluminium, leaving the devices free to actuate. We begin experiments when the aluminium is completely etched from beneath the particles, typically 12 hours after the sample is submerged. Experiments take place on the silicon chip.   

We visualize the device dynamics by optical microscope with a 10x objective. The microscope is integrated with a set of three Helmholtz coils, which provide uniform magnetic fields in three orthogonal directions. We used custom Python code to control the magnetic field direction and amplitude, as well as to provide real-time recordings of the magnetic field amplitude. The maximum magnetic field amplitude that can be generated by each Helmholtz coil is 44 G (4.4 mT), which is much smaller than the coercive fields of the nanomagnets (see Section ``Nanomagnet characterization" in SI). The particles thus act as if they have a permanent magnetization for the duration of the experiment. 

A Basler ac1300-60uc was used to record experimental images. A custom Python script captured time-stamped recordings, such that the magnetic field and particle dynamics could be synchronized. Videos were recorded at 30 fps. Custom MATLAB code was written to analyze the experimental images. 

\subsection*{Detailed fabrication flow}
\label{sec:fab}

The particle fabrication process primarily consists of three functional layers, each separated and sandwiched by structural silicon oxide layers that form the main body of the particle. The first functional layer comprises nanomagnet patterns defined via electron beam lithography and metallized by cobalt. In the subsequent layer, we pattern a reflective metallic marker for particle tracking under an optical microscope. The top functional layer is the hinge layer, which is a spin-coated polymer (anti-reflective coating for photolithography, Brewer Science DUV 42P-6). These functional layers are interleaved with SiO$_2$ layers, providing structural support and integrity to the particle. After the fabrication is finished, we dice the wafer into small pieces, each piece containing one device design. Devices can then be magnetized and released under a microscope by submerging sample pieces in a dilute solution of tetramethyl ammonium hydroxide in water($\sim 0.1$\%), which etches the aluminium release layer. We outline the detailed fabrication flow below. All fabrication steps were performed at the Cornell NanoScale Science and Technology Facility. 

\begin{enumerate}
        \item Define and etch stepper alignment markers on the front side of the wafer.  
        \begin{enumerate}
            \item Spin-coat and bake positive photoresist (UV 210-0.6). 
            \item Expose using a DUV stepper (ASML PAS 550/300C), post-exposure bake and develop by exposure to developer (AZ MIF 726).
            \item Etch using a CH$_4$ plasma for 5 minutes and 50 seconds in an Oxford PlasmaLab 80+ reactive ion etcher.
            \item Remove photoresist by oxygen plasma (5 minutes, Oxford PlasmaLab 80+).
        \end{enumerate}
        \item Define and etch electron-beam alignment markers on the front side of the wafer. 
            \begin{enumerate}
            \item Spin-coat and bake positive photoresist (UV 210-0.6). 
            \item Expose using a DUV stepper (ASML PAS 550/300C), post-exposure bake and develop by exposure to developer (AZ MIF 726).
            \item Etch using the Bosch process in a deep silicon etcher (Unaxis 770, 5 loops).
            \item Remove photoresist by oxygen plasma (5 minutes, Oxford PlasmaLab 80+).
        \end{enumerate} 
        \item Deposit and pattern the release layer. 
        \begin{enumerate}
            \item Deposit 200 nm aluminium by electron-beam evaporation (CVC SC4500). 
            \item To protect the release layer during subsequent processing, grow an additional 10 nm aluminium oxide by atomic layer deposition (Oxford ALD FlexAL). 
            \item Spin-coat and bake positive photoresist (UV 210-0.6). 
            \item Expose using a DUV stepper (ASML PAS 550/300C), post-exposure bake and develop by exposure to developer (AZ MIF 726).
            \item Etch aluminium by dry etching (7 minutes, PlasmaTherm 770)
           \item Remove photoresist by oxygen plasma (5 minutes, Oxford PlasmaLab 80+).
        \end{enumerate} 
        \item The first 200 nm layer of silicon oxide is deposited by high-density plasma CVD (Plasma-Therm Takachi HDP-CVD). 
        \item Define the nanomagnetic layer. 
        \begin{enumerate}
            \item Spin-coat and bake two layers of PMMA (495k PMMA A4 followed by 950k PMMA M2) to perform a bi-layer liftoff process. 
            \item Spin-coat an anti-charging agent (DisChem DisCharge H20x2). 
            \item Use electron-beam lithography (JEOL JBX-9500FS, 20 nA) to expose the resist. 
            \item After exposure, we rinse away the anti-charging agent in water, and develop by exposure to a 1:3 mixture of MIBK and IPA. 
            \item Descum the pattern by a mild oxygen plasma (3 seconds, Oxford PlasmaLab 80+). 
            \item Deposit 5 nm of titanium, followed by 50 nm of cobalt, and another 5 nm of titanium by electron-beam evaporation (CVC SC4500). The titanium serves as an adhesion and anti-oxidation layer for the cobalt. 
            \item Liftoff using Microposit Remover 1165 and under sonication.
        \end{enumerate}
        \item The second 100 nm layer of silicon oxide is deposited by high-density plasma CVD (Plasma-Therm Takachi HDP-CVD). 
        \item Define the marker layer by photolithography and etching
        \begin{enumerate}
            \item Sputter 30 nm of chromium (AJA sputter tool)
            \item Spin-coat and bake positive photoresist (DUV 210-0.6). 
            \item Expose the photoresist using a DUV stepper (ASML PAS 550/300C), post-exposure bake, and develop by exposure to developer (AZ MIF 726).
            \item Etch chromium by dry etching (5 minutes, PlasmaTherm 770)
            \item Remove photoresist by oxygen plasma.
        \end{enumerate}
        \item Etch the silicon oxide to define the particle shape. 
        \begin{enumerate}
            \item Spin-coat and bake positive photoresist (DUV 210-0.6). 
            \item Expose the photoresist using a DUV stepper (ASML PAS 550/300C), post-exposure bake, and develop by exposure to developer (AZ MIF 726).
            \item Etch through both silicon oxide layers in a CHF$_3$/O$_2$ plasma for 17 minutes (Oxford PlasmaLab 80+).
            \item Remove photoresist by oxygen plasma (5 minutes, Oxford PlasmaLab 80+). 
        \end{enumerate}
        \item Define the hinge layer
            \begin{enumerate}
            \item Spin-coat and bake 60 nm of anti-reflective coating (DUV 42P-6, this is the structural material for the hinge)
            \item Spin-coat and bake positive photoresist (DUV 210-0.6). 
            \item Expose the photoresist using a DUV stepper (ASML PAS 550/300C), post-exposure bake, and develop by exposure to developer (AZ MIF 726).
            \item Etch hinge material (1 min 50 seconds, Oxford PlasmaLab 80+). 
            \item Remove photoresist by organic solvents (Microposit remover 1165).
            \item Thin down the hinges by etching (30 seconds, Oxford PlasmaLab 80+). 
        \end{enumerate}
        \item Dice the sample into pieces (DISCO Dicing Saw). 
\end{enumerate}

\section*{Acknowledgements}
This work was performed in part at the Cornell Nanoscale Facility, an NNCI member supported by the National Science Foundation with Grant No. NNCI-2025233. This work made use of the Cornell Center for Materials Research shared instrumentation facility. This work made use of a Quantum Design MPMS-3 supported by NSF (DMR-1920086). The funding that supported this work includes NSF Designing Materials to Revolutionize and Engineer our Future, grant DMREF-89228, Alfred P. Sloan Foundation under grant No. G-2021-14198, and Cornell Center for Materials Research under grant No. DMR-1719875. J.Z.K. was supported by postdoctoral fellowships from Bethe/KIC/Wilkins, Mong Neurotech, and the Eric and Wendy Schmidt AI in Science program of Schmidt Sciences.

\section*{Data Availability Statement}
The data that support the findings of this article are openly available at~\cite{DataAvailability}. 

%\bibliography{references}

\end{document}